\documentclass{ptephy_v1}

\preprintnumber{XXXX-XXXX} 

\usepackage{amsmath} 
\usepackage{graphics} 
\usepackage{subfig} 
\usepackage{wasysym}
\usepackage[left]{lineno}

\definecolor{mygreen}{rgb}{0,0.4,0}

\begin{document}

\title{Development of Dual-phase Xenon TPC with a Quartz Chamber for Direct Dark Matter Search}


\author[1,*]{Kazufumi Sato\footnote{Present address: Institute for Space-Earth Environmental Research, Nagoya University, Nagoya, Aichi 464-8601, Japan}}
\author[1,2,*]{Masaki Yamashita$^\dag$}
\author[1,2]{Koichi Ichimura\footnote{Present address: Research Center for Neutrino Science, Tohoku University, Sendai 980-8578, Japan}}
\author[3,4]{Yoshitaka Itow}
\author[4,5]{Shingo Kazama}
\author[1,2]{Shigetaka Moriyama}
\author[3]{Kosuke Ozaki}
\author[1]{Takumi Suzuki}
\author[3]{Rina Yamazaki}
\affil[1]{Kamioka Observatory, Institute for Cosmic Ray Research, the University of Tokyo, Higashi-Mozumi, Kamioka, Hida, Gifu, 506-1205, Japan}
\affil[2]{Kavli Institute for the Physics and Mathematics of the Universe (WPI), the University of Tokyo, Kashiwa, Chiba, 277-8582, Japan}
\affil[3]{Institute for Space-Earth Environmental Research, Nagoya University, Nagoya, Aichi 464-8601, Japan}
\affil[4]{Kobayashi-Maskawa Institute for the Origin of Particles and the Universe, Nagoya University, Furo-cho, Chikusa-ku, Nagoya, Aichi, 464-8602, Japan}
\affil[5]{Institute for Advanced Research, Nagoya University, Furo-cho, Chikusa-ku, Nagoya, Aichi, 464-8602, Japan
\email{kazufumi@isee.nagoya-u.ac.jp, yamashita@isee.nagoya-u.ac.jp}}

\begin{abstract}
The idea of a hermetic quartz chamber in a dual-phase xenon time projection chamber (TPC) has the potential to improve the detector sensitivity for direct dark matter searches in the future. A major challenge facing TPC detectors in future dark matter experiments will be the reduction of the internal background such as $^{222}$Rn and the deterioration of the ionization signal due to electronegative impurities. The hermetic quartz chamber can isolate the TPC's sensitive volume from external interference and is thus expected to prevent contamination caused by the radioactive and electronegative impurities, which originate from outer detector materials. At the Kamioka Observatory in Japan, we developed a TPC with a quartz chamber that contains a $\diameter 48 \times 58$ mm volume of liquid xenon. At this development stage, we did not aim at the perfect hermeticity of the quartz chamber. Our aim here is twofold: first, to demonstrate via the use of a calibration source that the presence of quartz materials in the TPC does not impact its operation; and second, to perform quantitative measurements of the TPC's characteristics. We successfully measured electron drift velocities of 1.2--1.7 mm/$\mu$s in liquid xenon under electric fields ranging from 75--384 V/cm, and also observed small S2 signals produced by a single ionized electron with a light yield of 16.5 $\pm$ 0.5 PE. These results were consistent with the expected values; therefore, our demonstrations provide a proof of principle for TPCs incorporating a quartz chamber.
\end{abstract}

\subjectindex{dark matter, liquid xenon, low background}

\maketitle

\section{Introduction}
Weakly interacting massive particles (WIMPs) are hypothetical particles thought to be present in the universe as cold dark matter. Experiments in underground laboratories that aim to observe elastic scattering of WIMPs directly using atomic target nuclei are either ongoing or have been proposed \cite{PDG2018}. Among these experiments, those using liquid noble gas (e.g., liquid xenon (LXe) and liquid argon (LAr)) experiments are leading the field, by providing the most sensitive WIMP-nucleus elastic cross-sections (see \cite{XMASSFV, LZ, XENONnT, DS-20K}). The sensitivity of direct searches for nuclear recoil signals at low energy ($<$ 100 keV) is often limited by the radioactive background caused by $^{222}$Rn emanation from detector materials, as well as by the energy threshold of the detectors. The successful identification of nuclear recoil signals from an electron recoil caused by background $\gamma$ and/or $\beta$ rays, and the use of a large target mass in the detectors, are essential for the search. In this regard, time projection chambers (TPCs) containing dual-phase Xe or Ar (i.e., consisting of liquid and gas) have been proven to be successful and have the potential to achieve WIMP--nucleus spin-independent elastic cross-sections of $10^{-48}$ cm$^2$ or smaller in the future \cite{DARWIN}.
For the next generation of dark matter experiments, it is crucial to further reduce the radioactive $^{222}$Rn contamination to $<$ 1 $\mu$Bq/kg; at this level, the contamination becomes comparable to the event rate of pp solar neutrinos at a low energy, thereby making it possible to obtain WIMP-nucleon cross-sections $\leq 10^{-48}$ cm$^2$.  
 
One way of addressing this problem is to use a hermetic TPC whose sensitive volume is separated from external components such as photosensors, electric circuits, and cables to avoid Rn emanation from the materials of these components. Quartz is a suitable material to separate volumes and make a hermetic TPC since it is known to have a small concentration of radium~\cite{XMASSPMT}. Therefore, it emits a negligible amount of $^{222}$Rn.
A hermetic TPC made of quartz is also promising for low-mass WIMP searches using the so-called S2-only analysis~\cite{XENON10, XENON100, DS50S2ONLY,XENON1TS2ONLY}. Owing to the large bandgap of quartz 9 eV, it is possible to avoid the secondary electron emission caused by the VUV scintillation of LXe (7 eV) \cite{Vella}. 
There are further advantages as follows:
(1) In future detectors in general, the drift length of the TPC is proposed to be approximately 1.5 m or more, which requires the ionized-electron lifetime to be 1 ms or more to avoid a charge signal deficit. A hermetic TPC made of quartz is useful for the reduction of impurities that capture ionized-electrons. 
(2) As the transparency of quartz under an LXe VUV light source is high ($\sim$ 90\%), scintillation can be detected directly without using a wavelength shifter.
(3) Finally, the quartz chamber prevents drift electrons generated in gas from entering the sensitive volume. These electrons produce noise events with small S2 signals and act as a background for sub-GeV dark matter searches~\cite{XENON100,XENON1TS2ONLY}.
Thus, a quartz chamber facilitates the search for WIMPs with a GeV--TeV mass, as well as other dark matter candidates with sub-GeV masses.

Though there has been no development of a hermetic TPC aiming to minimize the level of radioactive $^{222}$Rn or realize a low mass WIMP search via a S2-only analysis, there were an experimental study and proposal using a hermetic TPC in the past. A hermetic TPC was constructed for the XMASS R\&D program for dark matter exploration by the Waseda group at Kamioka in the early 2000s \cite{Yamashita, XMASS}. Gas and liquid xenon were surrounded by PTFE cones and two MgF$_2$ windows in a stainless-steel or copper vessel. Two photomultipliers (PMTs) were coupled to these windows through a 1-mm-thick vacuum. The hermetic structure protected the PMTs from undergoing immersion in LXe, thereby satisfying the requirement for low temperature (-100 $^\circ$C) operation. In addition, the hermetic structure helps to mitigate the deterioration of detector performance caused by the unavoidable loss of scintillation at the intersection of the optical window and vacuum. The Waseda group, however, encountered difficulty when holding the pressure against the vacuum whilst scaling up to a $\gtrsim$ 1 m size optical window. However, it is notable that the detector achieved an electron lifetime of 3 ms without the introduction of the online purification of xenon gas during the operation.
Another hermetic design using dual-phase Xe was proposed by the XAX experiment in 2009 as a multi-purpose detector searching for dark matter, low energy solar neutrinos, and neutrino-less double beta decay ~\cite{XAX}. The objective of the experiment was to produce a multi-ton, multi-target detection system incorporating a hermetic acrylic vessel with a wavelength shifter inside the wall holding the LXe or LAr target. Interestingly, the possibility of using ultrapure LXe for such multi-purpose physics targets had initially been discussed prior to the XAX experiment, for a 10-ton single-phase detector (see~\cite{SUZUKI}). 
 
Despite its obvious advantages, quartz is known as an insulator and accumulate charges, which potentially poses a problem during TPC operation as the electric charge is altered. We therefore proposed a new design for a dual-phase Xe TPC including a hermetic quartz chamber, with a prototype built and tested at the Kamioka Observatory, Japan. 
In this paper, we report the basic properties of the prototype TPC with the quartz chamber, including quantitative measurements of the scintillation light yield, the drift velocity of ionized electrons, and single electron gain. As we explain later, the quartz chamber in our prototype is not entirely sealed since our design uses small holes to exchange gas/liquid for the gas purification. The demonstration of the protection performance against $^{222}$Rn and electronegative impurities will be the subject of future work. 
Section 2 describes the detector apparatus, while Section 3 details the signal measurement methodology. The results and discussion are presented in Sections 4 and 5, respectively. 

\section{Apparatus}
\begin{figure}[!h]
\centering
\includegraphics[width=0.98\textwidth]{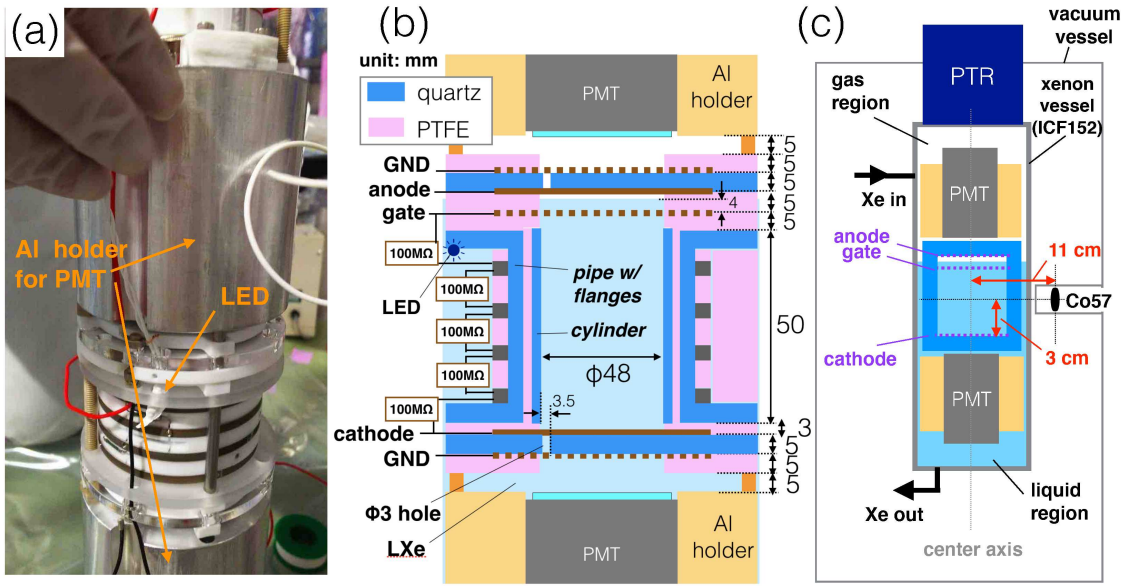}
\caption{Setup of a prototype of TPC. (a) Photograph of TPC. (b) Schematic view of the TPC cross-section. (c) Schematic view of LXe and vacuum vessels.}
\label{fig:detector}
\end{figure}

The prototype TPC is shown in Fig.~\ref{fig:detector}(a). 
The TPC consists of a quartz chamber, electrodes, and two PMTs. The PMTs are covered with aluminum holders in the picture. Figure~\ref{fig:detector}(b) shows a cross-section view of the TPC.
All the quartz materials used in the TPC are ES-grade (Tosoh Quartz Corporation), which, for a 5 mm thickness, has 92\% transparency for LXe scintillation light wavelength ($\lambda \sim$ 175~nm \cite{FUJI}). A $\phi$95~mm $\times$ 5~mm plate is mounted on the top and bottom of the chamber. Each plate has a $\phi3$~mm hole so that Xe can be circulated to purify the inner volume. 
A 50 mm length pipe with $\phi$95 mm flanges is placed in the middle. The thickness of the pipe is 5~mm, and the inner diameter is $\phi$75 mm. 
In the innermost part of the quartz chamber, there is a cylinder with $\phi$48 mm inner diameter, 2.5~mm thickness, and 50 mm length.  
PTFE support structures are used between these quartz components.  
 Anode and cathode electrodes are attached to the inner surfaces of the top and bottom quartz plates, respectively. These electrodes are chemically etched square grids made from 0.1 mm thick stainless-steel. The grid's spacing and width are 2 mm and 0.1 mm, respectively. The distance between the anode and the cathode electrode is 63 mm.
A gate electrode made of $\phi0.1$ mm gold-plated stainless-steel wires arranged in parallel with 2 mm spacing is placed 5 mm below the anode electrode. 
High voltages of +2250, $-2250$, and $-2750$ V are typically applied to the anode, gate, and cathode electrodes, respectively, to create a downward electric field. 
Four stainless-steel field-shaping rings with a $5\times 5$~mm$^2$ square cross-section are spaced equally between the gate and cathode electrodes to produce a uniform electric field in the TPC.
Each of them, including the gate and cathode electrodes, are connected in series with 100~M$\Omega$ resistors.
Two grounded electrodes with stainless-steel wires are attached to the outer surfaces of the top and bottom quartz plates to prevent the PMTs from being affected by the electric field.
The chamber is filled with LXe, and the liquid level is adjusted to be 1 mm below the anode. The level is monitored by a capacitance level sensor mounted outside the quartz vessel. 

When charged particles deposit energies in LXe, scintillation photons (S1) are produced via de-excitation or recombination processes~\cite{XENONPROCESS}. A proportion of the electrons generated by the ionization process drift upward due to the cathode--gate electric field and subsequently produce secondary scintillation photons (S2) owing to the high gate--anode electric field. In order to detect S1 and S2 scintillations, two PMTs (HAMAMATSU R10789) are placed below and above the top and bottom quartz plates, respectively. They are covered with aluminum holders not only to support the electrode structure but also to save the amount of LXe. 
The top PMT signal is fed into the $\times 10$ amplifier (Phillips model 776) and is recorded by a 1GHz sampling 10-bit ADC waveform digitizer with a 1 V$_{pp}$ input dynamic range (CAEN v1751). The bottom PMT signal is split into two, with only one of these signals fed into the amplifier. Both signals are recorded with the same digitizer as the top PMT.
Data acquisition is triggered by the coincidence of the top and (amplified) bottom PMT S1 signals, and waveforms ranging from $-8$--$192$ $\mu$s with respect to the trigger timing are stored for every event.  

The TPC is placed inside a vessel containing LXe, which consists of a stainless-steel 304 ICF152-standard tube, as shown in Fig.~\ref{fig:detector}(c). The upper side of the LXe vessel is connected to a pulse tube refrigerator (PTR) to maintain the operating temperature around at $-100.0\pm0.3$ $^\circ$C, and the gauge pressure is typically 0.0715 $\pm$ 0.0015 MPa. The outer vessel is kept in a vacuum for thermal insulation. Owing to the presence of electronegative impurities (e.g., O$_2$), a number of drift electrons are absorbed before reaching the gate electrode, and the S2 signal is strongly attenuated. Therefore, a recirculation system is used to purify Xe gas. The LXe vessel has two gas ports to fill and extract Xe gas, and the gas is recirculated by a metal bellows pump. The extracted gas is passed at a flow speed of 10 SLPM through a getter purifier (API-Getter, NIPPON API Co. LTD.) to remove impurities and is then returned into the LXe vessel. A heat exchanger \cite{HEATEX} is used at the inlet (gas) and outlet (liquid) points to save the cooling power and achieve stable conditions. 

To study the base performance of the TPC, we measured S1 and S2 signals, which originate from the 122 keV $\gamma$ rays of the $^{57}$Co calibration source. The vacuum vessel has a tube that houses an 80 kBq $^{57}$Co source. The source is placed at a distance of 11 cm from the TPC center in a radial direction and 3 cm above the cathode electrode, from where it irradiates 122 keV $\gamma$ rays toward the entire of the TPC as shown in Fig~\ref{fig:detector}(c).

\section{Analysis}
\subsection{Definition of S1 and S2 signals}
\label{sec:definition-s1-s2}
\begin{figure}[!h]
\centering
\includegraphics[width=0.85\textwidth]{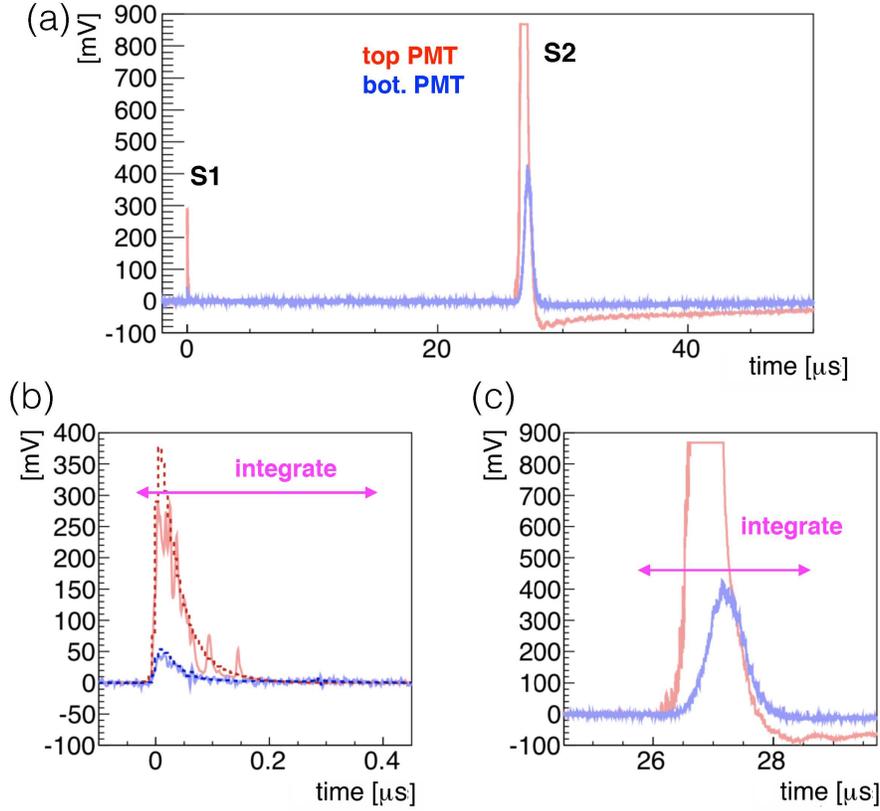}
\caption{Waveforms from the top PMT (light red) and bottom PMT (with the amplifier, light blue) recorded using a waveform digitizer. (a) Overall view. (b) Enlarged view around the S1 signals. The average waveforms of the top and bottom PMTs are indicated by dark-red and dark-blue dashed lines, respectively. (c) Enlarged view around the S2 signals. The flat-top shape of the top PMT signal is due to the saturation of the ADC.}
\label{fig:waveform}
\end{figure}

Figure~\ref{fig:waveform} shows the waveforms typically observed for the top and bottom PMTs. The top PMT operated at a higher voltage (1275 V) to observe single photoelectrons (PE), while the bottom PMT operated at 1000 V to avoid the S2 signal saturating the ADC. The PMT gains are about $1.8 \times 10^6$ and $1.2 \times 10^5$ for the top and bottom PMT, respectively, before being fed into the amplifier. The PMT gains were measured by using a single PE produced by a blue LED, which was installed next to the outer wall of the quartz chamber.

S1 pulse timing was determined by the constant fraction method, i.e., the point at which the waveform crosses half of the peak height. We evaluated the S1 charge $PE_{S1}$ by integrating a waveform in the -20--400 ns range with respect to the S1 pulse timing and converting it to the number of PE. For the S2 signal, we searched for the peak exceeding 10 mV ($\sim 70$ PE) in the waveform of the non-amplified bottom PMT signal. The S2 signal charge $PE_{S2}$ was calculated by integrating within $\pm$ 1.5 $\mu$s from the peak. The S2 timing $t_{S2}$ was determined as a weighted mean within the integral range. Then, the drift time of S2 $t_{drift}$ was determined as $t_{drift} \equiv t_{S2} - t_{S1}$. 

\begin{figure}[!h]
\centering
\includegraphics[width=0.98\textwidth]{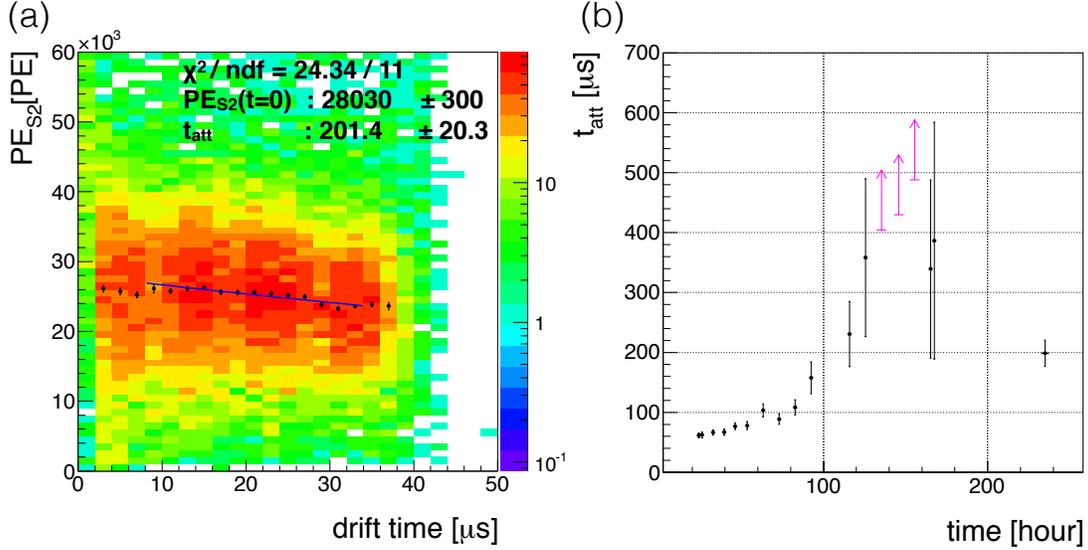}

\caption{(a) S2 signal attenuation as a function of drift time after a 237-hour circulation. The black points with error bars show the peak value of each time bin. The solid line shows the fit for the data described by Eq.~(\ref{driftLength}). (b) Electron lifetime $t_{att}$ as a function of circulation time. The error bars show the statistical errors associated with data fitting. The up-arrows indicate the lower limit of the electron lifetime.}
\label{fig:purification}
\end{figure}

Significant attenuation of the S2 signals was observed at the beginning of the test. This attenuation was recovered by recirculation, as shown in Fig.~\ref{fig:purification}(a). 
We evaluated the lifetime $t_{att}$ of the drift electron by fitting with the following equation:

\begin{equation}
\label{driftLength}
 PE_\mathrm{S2}(t_{drift}) = PE_\mathrm{S2}(t_{drift}=0)\times\exp\left(-\frac{t_{drift} }{t_{att}}\right).
\end{equation}

After 100 hours of recirculation, we achieved $t_{att} \gtrsim$ 200 $\mu$s as depicted in Fig.~\ref{fig:purification}(b). The following analysis was performed using the data collected after 160 hours of recirculation. The electron lifetime measurement in the period between 170 and 230 hours of recirculation is not shown here as the electron drift velocity measurement was performed in that period. The electron drift velocity measurement is described in  \ref{sec:drift-time-distr}.

\subsection{$^{57}$Co selection}

\begin{figure}[htb]
\centering
\includegraphics[width=\textwidth]{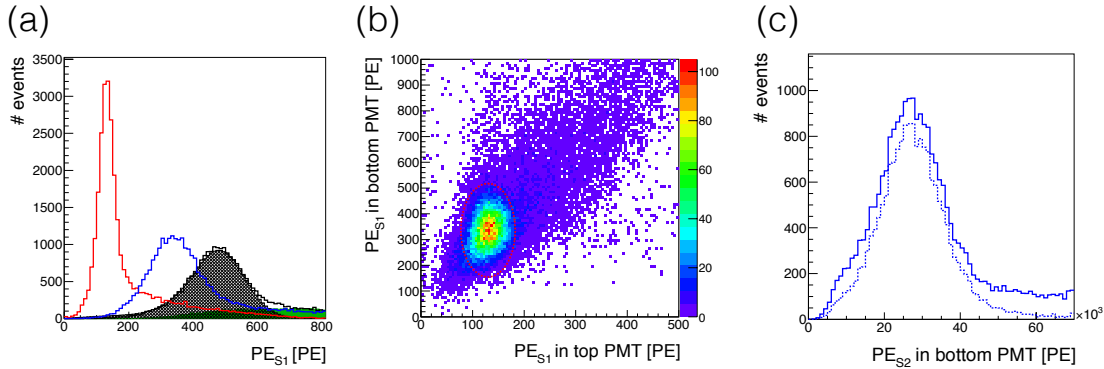}
\caption{(a) $PE_{S1}$ distribution observed in the top PMT (red), bottom PMT (blue), and the sum of both PMTs (black) together with the MC expectation (black-hatched). Background data, i.e. without $^{57}$Co source, is also shown (green-filled). (b) Correlation of $PE_{S1}$ between the top and bottom PMTs. The red dashed line shows the selected region based on S1 in Eq.~(\ref{eq:1}). (c) $PE_{S2}$ distribution observed in the bottom PMT. The solid (dashed) histogram shows the distribution without (with) the $^{57}$Co S2 selections.}
\label{fig:s1spectrum}
\end{figure}

Figure~\ref{fig:s1spectrum}(a) shows the observed $PE_{S1}$ distribution for the top and bottom PMTs, as well as the sum from both. In the case of the summed signal, the background data recorded without the $^{57}$Co source are also shown (green).  The summed signal exhibits a clear peak that is attributed to the 122 keV $\gamma$ rays originating from the $^{57}$Co source. 
 To understand the detector performance, we performed a full detector Monte Carlo (MC) simulation using the Geant4 package \cite{GEANT4}. The simulation accounted for the particle tracks, scintillation process, propagation of scintillation photons, and PMT response based on the XMASS MC package \cite{XMASSFV}. Several optical parameters were used for the LXe scintillation light ($\sim$ 175 nm) in the MC simulation; these included the following: the absorption (5 m) and scattering coefficients (52 cm) of the LXe scintillation light, the reflectance at the PTFE reflector (95\%), the refractive indices of LXe (1.62) and quartz (1.59), and the reflection and absorption probabilities at a PMT photocathode. The quantum efficiency of the PMTs at 175 nm were provided by the manufacturer: 31.1\% and 31.3\% for the top and bottom PMTs, respectively. The result of the MC simulation is depicted in Fig~\ref{fig:s1spectrum}(a) for the total number of PE for the irradiation of 122 keV $\gamma$ rays. We found that the MC prediction agrees well with the experimental result. For each PMT, the difference between the measured data and the MC simulation was approximately 10-20\%.

To select the $^{57}$Co events, we required 
\begin{equation}
\label{eq:1}
 \sqrt{\left(\frac{PE_{S1}^{top}-\langle PE_{S1}^{top}\rangle}{\sigma^{top}}\right)^2 + \left(\frac{PE_{S1}^{bot}-\langle PE_{S1}^{bot}\rangle}{\sigma^{bot}}\right)^2}<2,
\end{equation}
where the superscript top (bot) indicates measurement in the top (bottom) PMT, and $\langle PE_{S1}\rangle$ and $\sigma$ are the mean and width of the $PE_{S1}$ distribution evaluated using a Gaussian fit, respectively. The selected region is illustrated in Fig.~\ref{fig:s1spectrum}(b).
  
To select an S2 signal from $^{57}$Co events, we applied the following criteria: the pulse width (FWHM) had to be $> 100$ ns to distinguish it from the S1 pulse, and the drift time $t_{drift}$ had to be $>$~1~$\mu$s to avoid the tail of S1 pulses. To eliminate multi-site events in the detector, such as Compton scattering, we chose only those events that contained a single S2 signal. Figure~\ref{fig:s1spectrum}(c) shows the S2 PE distributions before and after $^{57}$Co selection. 

\section{Results}

\subsection{Electron drift velocity measurement}
\label{sec:drift-time-distr}
\begin{figure}[htb]
\centering
\includegraphics[width=\textwidth]{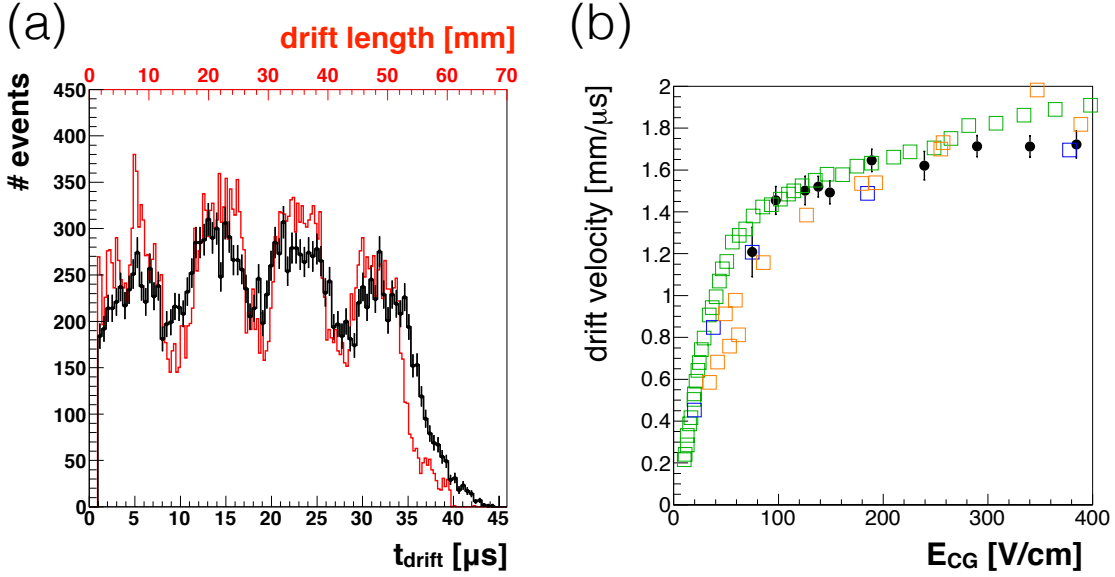}
\caption{(a) Drift time $t_{drift}$ distribution. The black and red histograms show the experimental data and the MC expectation assuming the drift velocity to be 1.52$\pm$ 0.050 mm/$\mu$s, respectively. The divisions on the top x-axis indicate the drift length. (b) Drift velocity as a function of the electric field. Our measurements are represented by filled black circles, with the colored squares showing the measurements reported by Miller et al. (orange\cite{MILLER}), Gushchin et al. (green \cite{GUSH}), and EXO-200 (blue \cite{ALBERT}). }
\label{fig:dtdist}
 \end{figure}

 To demonstrate that the quartz chamber works properly as a TPC, we measured the electron drift velocity within the LXe. 
 For the $^{57}$Co data, four peaks were observed in the $t_{drift}$ distribution as shown in Fig.~\ref{fig:dtdist}(a). The data were taken with the electric field between the cathode and gate ($E_{CG}$) measuring 138 V/cm. The average electric field along the drift path was estimated using COMSOL Multiphysics software~\cite{COMSOL}. Based on the results of the MC modeling, we confirmed that these four peaks can be attributed to the shadow effect of the field-shaping rings on the $\gamma$ rays produced by the $^{57}Co$ source; the modeling further enabled us to estimate the drift velocity of ionized electrons in LXe using the linear relation between $t_{drift}$ and the location of the peaks. Thus, from the data presented in Fig~\ref{fig:dtdist}(a), we evaluated the drift velocity to be 1.52 $\pm$ 0.050~mm/$\mu$s at an electric field of 138 V/cm. The error of the measurement includes both the statistical error and the error arising from repeatability. 
We acquired additional drift velocity data by changing $E_{CG}$ from 75--384 V/cm. As Fig.~\ref{fig:dtdist}(b) illustrates, the electron drift velocity within this electric field range was measured from 1.2--1.7 mm/$\mu$s, which is consistent with the measurements from other studies~\cite{MILLER, GUSH, ALBERT}. Our measurements were performed over a period of 60 hours following 170 hours of Xe gas recirculation; therefore, we concluded that no charge-up effects were observed on such a time scale. 

\subsection{Single electron sensitivity and S2 charge yield}
\label{sec:single-electr-sens}
\begin{figure}[t]
\centering
\includegraphics[width=0.98\textwidth]{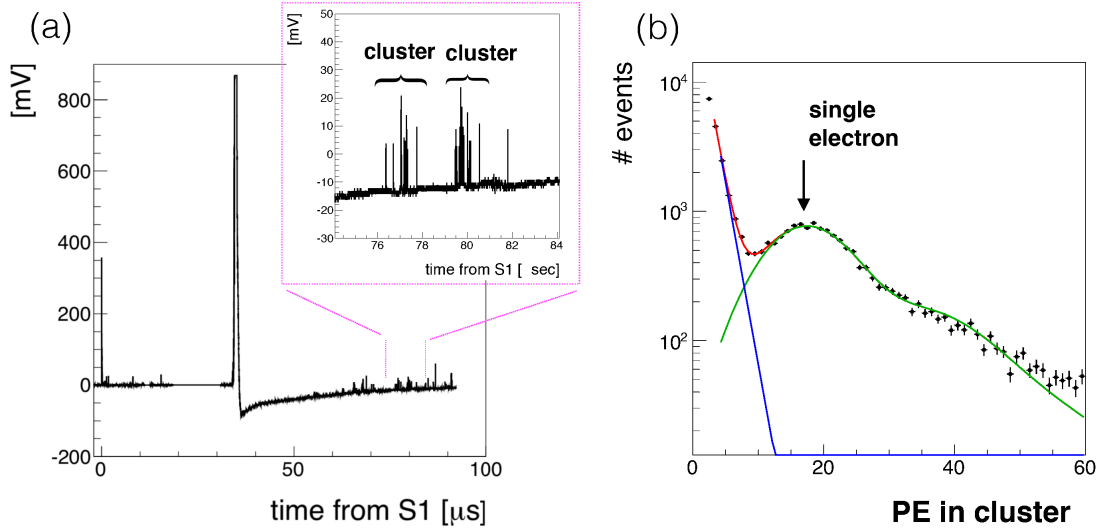}
\caption{(a) Cluster observed in the waveform, which is a candidate for the S2 signal from a single drift electron. In this event, we found four clusters with $> 10$ PE. An enlarged view of the region of 75--84 $\mu$s is depicted in the right upper box. (b) PE distribution of the S2 signal from the top PMT. The green line shows a convolution of the Gaussian and Poisson distribution, and the blue line shows an exponential function. The red line shows the sum of the green and blue functions, which was used as a fitting function to evaluate the single drift electron peak.}
\label{fig:singleElectron}
\end{figure}

Sensitivity of the detector to small S2 signals is a key factor in the exploration of sub-GeV dark matter using the S2-only analysis. Therefore, we investigated the S2 signal for an extremely low energy region (i.e., the S2 signal produced by a single ionized electron). LXe scintillation photons have an energy of 7 eV ($\lambda$ $\sim$ 175 nm), which is sufficient for the production of single electrons via the photoelectric effect on the detector material surface or via photo-ionization with impurities in LXe (e.g., $O_2^-$) \cite{Sorensen}. Therefore, small S2 signals originating from single electrons are expected to appear after the large $^{57}$Co S2 pulse, and can be used for the calibration to demonstrate this capability. 
 In fact, we observed clustered PE peaks after the $^{57}$Co S2 signal as shown in Fig.~\ref{fig:singleElectron}(a). The cluster has a standard deviation of approximately a 170 ns time spread, which is consistent with the pulse width of the $^{57}$Co S2 signal at $t_{drift}=0$. The clusters only appear within $\sim$45 $\mu$s after $^{57}$Co S2 signal, which corresponds to the maximum drift time of $^{57}$Co S2 shown in Fig.~\ref{fig:dtdist}(a). From the observed width and timing from $^{57}$Co S2 signal, we concluded that those clusters are not S1 or accidental backgrounds. We can also reject the possibility that the clusters come from PMT after-pulses, which occur within $<$ 1 $\mu$s after the large PMT pulse. 
 The origin of the small number of PE pulses below 10 PE is uncertain. However, this phenomenon was also observed in other studies~\cite{LUX1PE, Sorensen}, which discussed the possibility of photon emission from PTFE as a result of the LXe VUV scintillation light. In fact, this component drops exponentially at below 10 PE, as shown in Fig.~\ref{fig:singleElectron}(a), hence it does not affect the measurement of the single electrons.
We evaluated the PE distribution in the 30--50 $\mu$s region from the $^{57}$Co S2 pulse to avoid the undershoot region and found that for single drift electrons a peak occurs at $ 16.5 \pm$ 0.5 PE for the top PMT, as shown in Fig.~\ref{fig:singleElectron}(b). The error includes the statistical error associated with data fitting, and the systematic error arising from different drift fields $E_{CG}$.  
The expected number of PE, $N_{PE}^{exp}$, produced by the single electron was calculated as
\begin{equation}
 N_{PE}^{exp} = Acc\left(\frac{dN_{ph}}{dx_{gas}}\cdot x_{gas}\right),
\end{equation}
where $Acc = 6.8\%$ is the acceptance of S2 photons detected at the top PMT evaluated by the MC simulation, and $x_{gas} = 1 \pm 0.6$ mm is a luminescence region in the gas. The $\frac{dN_{ph}}{dx_{gas}}$ is an S2 photon yield per unit gas thickness obtained from the empirical formula given in \cite{CHEPEL}:
\begin{equation}
\frac{dN_{ph}}{dx_{gas}} = 0.137 E_{gas}\mathrm{[V/cm]}-177P\mathrm{[bar]}-45.7,
\end{equation}
where $E_{gas} = 14.5 \pm 0.1$ kV/cm and $P = 1.715 \pm 0.015$ bar are the electric field and absolute pressure in the gas, respectively. We estimated $N_{PE}^{exp}$ to be 12.7 $_{-5.5} ^{+ 4.4}$ PE. The direct measurements of the number of PE produced by the single electron are consistent with this expectation value. For comparison, in the case of the bottom PMT, $N_{PE}^{exp}$ was estimated as 10.9 $\pm$ 0.33 PE per single electron due to the geometrical difference relative to the top PMT.

\begin{figure}[!ht]
\centering
\includegraphics[width=0.48\textwidth]{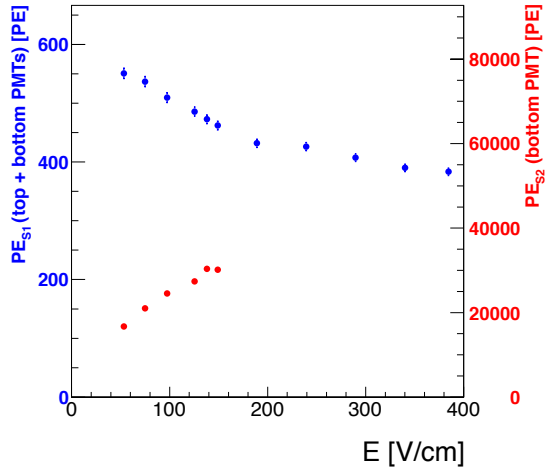}
\caption{$PE_{S1}$ (top and bottom PMTs)(blue) and $PE_{S2}$(bottom PMT) (red) of $^{57}$Co events as a function of $E_{CG}$. The electron lifetime was considered to correct the charge loss during the drift for the S2 signal.} 
\label{fig:driftFieldDep}
\end{figure}

Finally, we cross-checked the charge yield of the S2 signal from the 122 keV $\gamma$ rays using the number of PE per single electron for the S2 signal. Figure~\ref{fig:driftFieldDep} shows the S1 and S2 PE yield dependency as a function of the electric field. The electron lifetime was considered to correct the charge loss during the drift for the S2 signal yield.
For instance, the S2 PE yield at 138 V/cm after lifetime correction was 3.1 $\times10^4$ PE, which changed to 2.8$\times 10^3$ electrons when the gain factor of 10.9 PE/electron for the bottom PMT was used. We estimated the expected charge yield of electrons for 122 keV $\gamma$ rays from the results presented in Fig.3 in \cite{APRILE}. To evaluate the total charge yield ($Q_0$), we used the relation $Q_0 = E_e/W_e$, where $E_e$ is the energy (i.e. 122 keV), and $W_e$ = 15.6 eV \cite{TAKAHASHI} is the average energy required to produce an electron-ion pair in LXe.  
Considering our electric field and its associated uncertainty, the estimated charge yield is (3.0 $ \pm$ 0.3) $\times 10^3$, which agrees well with the quantitative measurement of the charge yield.

%
%
%
%
%

\section{Conclusion and Discussion}

The reduction of radioactive $^{222}$Rn and electronegative impurities emanated from detector materials is a critical issue that must be addressed for the improvement of future direct dark matter searches using Xe TPCs. The use of the hermetic quartz chamber is proposed as a countermeasure to resolve this issue.

We developed a small prototype for a TPC with a quartz chamber and demonstrated its successful operation. We measured drift electron velocities of 1.2--1.7 mm/$\mu$s in LXe under electric fields ranging from 75--384 V/cm. The obtained results were consistent with measurements reported by other studies. We ran the quartz chamber for more than 200 hours in this study, and the chamber was found to perform effectively without the problem of charge-up on the TPC on such a time scale. Furthermore, we showed that the TPC could detect small S2 signals produced by a single drift electron with a light yield of 16.5 $\pm$ 0.5 PE, and it is found that the S2 charge yield is in good agreement with the value reported by other experimental studies.

 We therefore conclude that our proposed TPC with the quartz chamber works successfully as an Xe TPC. This study serves as a basis for further R\&D to test the reduction of $^{222}$Rn and electronegative impurities. Potential problems related to the uniformity of the electric field with a large detector and the operational stability for run times exceeding 240 hours are the topics for future studies. 

\section*{Acknowledgment}
We are grateful for the cooperation of the Kamioka Mining and Smelting Company. This work was supported by the Japanese Ministry of Education, Culture, Sports, Science and Technology, Grant-in-Aid for Scientific Research, JSPS KAKENHI Grant No. 17H05197, 19H05805, and 26104004, and the joint research program of the Institute for Cosmic Ray Research (ICRR), the University of Tokyo. 


%

\end{document}